\documentclass[aps,floats,superscriptaddress,twocolumn,pre,showpacs]{revtex4-1}

\usepackage{graphicx,epsfig}
\usepackage{graphics,dcolumn,bm,epic, eepic,fleqn,float}
\usepackage{amssymb,amsmath,amsfonts,multirow,rotate,color}
\usepackage{soul}
\usepackage{wasysym}
\usepackage{dcolumn}
\usepackage{bm}

\newcommand{\lay}[1]{^{[#1]}}
\def\Erdos{Erd\"os}

\begin{document}

\title[Layered social influence promotes multiculturality in the Axelrod model]{Layered social influence promotes multiculturality in the Axelrod model}

\author{Federico Battiston}
\affiliation{School of Mathematical Sciences, Queen Mary University of London, London E1 4NS, United Kingdom}

\author{Vincenzo Nicosia}
\affiliation{School of Mathematical Sciences, Queen Mary University of London, 
London E1 4NS, United Kingdom}

\author{Vito Latora}
\affiliation{School of Mathematical Sciences, Queen Mary University of London, 
London E1 4NS, United Kingdom}

\author{Maxi San Miguel}
\affiliation{IFISC (CSIC-UIB), E-07071 Palma de Mallorca, Spain}



\begin{abstract}
Despite the presence of increasing pressure towards globalisation, the coexistence of different cultures is a distinctive feature of human societies. However, how multiculturality can emerge in a population of individuals inclined to imitation, and how it remains stable under cultural drift,  i.e. the spontaneous mutation of traits in the population, still needs to be understood. To solve such a problem, we propose here a microscopic model of culture dissemination which takes into account that, in real social systems, the interactions are organised in various layers corresponding to different interests or topics. We show that the addition of multiplexity in the modeling of our society generates qualitatively novel dynamical behavior, producing a new stable regime of cultural diversity. This finding suggests that the layered organisation of social influence typical of modern societies is the key ingredient to explain why and how multiculturality emerges and thrives in our world. 
\end{abstract}

\maketitle
%
%
\thispagestyle{empty}


\section*{Introduction}

Our understanding of societies, and social dynamics more generally, is increasingly taking advantage of concepts,
models and methods from statistical and computational
physics~\cite{oliveira99,castellano09}.
Within this framework, the
validitation of sociological theories relies on the construction of
appropriate mechanistic agent-based models~\cite{Dignum2014} which can be tested against
existing and upcoming large data sets.

The existence of multiculturality and group boundaries is a
well-established feature of social systems~\cite{fredik69, boyd05} and much effort has been devoted into explaining possible mechanisms able to reproduce such empirical finding.
In particular, cultural diversity appears as a striking phenomenon, supposedly in contrast with the widely acknowledged principle of social influence~\cite{festinger63}.
According to such mechanism, changes in the cultural
traits of an agent are influenced by the acquaintances and friends of
that agent facilitating the local
convergence towards a set of common cultural traits and promoting homogeneity across the populations.
For instance, homogenous absorbing states are typical of imitative dynamics where the state of the individuals is described through a scalar binary variable, as for the voter model~\cite{liggett85} in finite size populations.

An interesting solution to reconcile social influence and the emergence of cultural diversity (polarization) commonly observed
in human societies was suggested by Robert Axelrod, who proposed
in 1997 a simple agent-based mechanistic model of dissemination of cultural traits~\cite{axelrod97}.
In the original model the state of individuals is not described through a scalar variable,
but each agent is endowed with a set of cultural features $F$,
each of them taking one of a number of different
cultural traits $q$.  Based on pairwise interactions among agents, together with social influence the model mimics
another important social principle, known as homophily~\cite{homans62,mcpherson87, mcpherson01}, i.e., the tendency
of individuals to connect and interact preferentially with similar
ones. As a result of pairwise interactions, imitation still occurs but is limited to the update of one feature at a time.
Homophily and social influence acting together constitute a self-reinforcing dynamics leading to local homogenization. However, a main result of the analysis of Axelrod was that in spite of local convergence, global polarization was possible: agents become more similar by local interactions, but cleavages among different cultural groups are created so that these groups no longer interact. The overall result is the possible emergence of a globally polarized or multicultural state with coexistence of different cultural groups. A quantitative analysis of the model~\cite{castellano00} unveiled a non-equilibrium phase transition at a critical number of cultural traits $q_c$ from a mono-cultural phase (global culture) to a
polarized or multicultural phase, where several groups with different
cultural traits survive. Such phenomenon has been observed also for realistic interaction patterns, such as small-world networks and heterogeneous distributions of the connections among the individuals~\cite{klemm03sw}.

However, it was found that the multicultural phase is achievable only
for a high initial number of traits $q$ in the population. In addition, the multicultural phase is not
stable under cultural drift, meaning that the spontaneous tendency of
agents to modify their cultural traits independently of their
environment drives the system towards a monocultural
state~\cite{klemm03, klemm05}.
Different mechanisms have been invoked to account for robust cultural
diversity, such as the plasticity of the social relations among agents, so that the social networks coevolves with the
dynamics of the cultural state~\cite{centola07}. Other alternative proposals
focus on modifications of the form of the local interactions, for
instance based on assimilation-contrast theory~\cite{sherif61},
also known as bounded confidence~\cite{deffuant00,weisbuch04, desanctis09}. It has been shown how by integrating metric cultural states~\cite{valori11,stivala14,babeanu16}, as opposed to nominal states, the bounded confidence mechanism can lead to robust cultural diversity~\cite{Flache2006}. 
Another type of modification~\cite{Flache2011} addresses the nature of social influence in line with other models of social contagion~\cite{Thr,watts,contagion,eguiluz}: the dyadic interpersonal influence of the original model is replaced by a mechanism in which an agent modifies its cultural state depending on the state of all agents in her neighborhood. Finally, the mechanism of social differentiation in which agents tend to increase cultural differences has also been analyzed in~\cite{Flache2011b,Mas2014}.

\begin{figure*}[t]
\begin{center}
\includegraphics[width=6in]{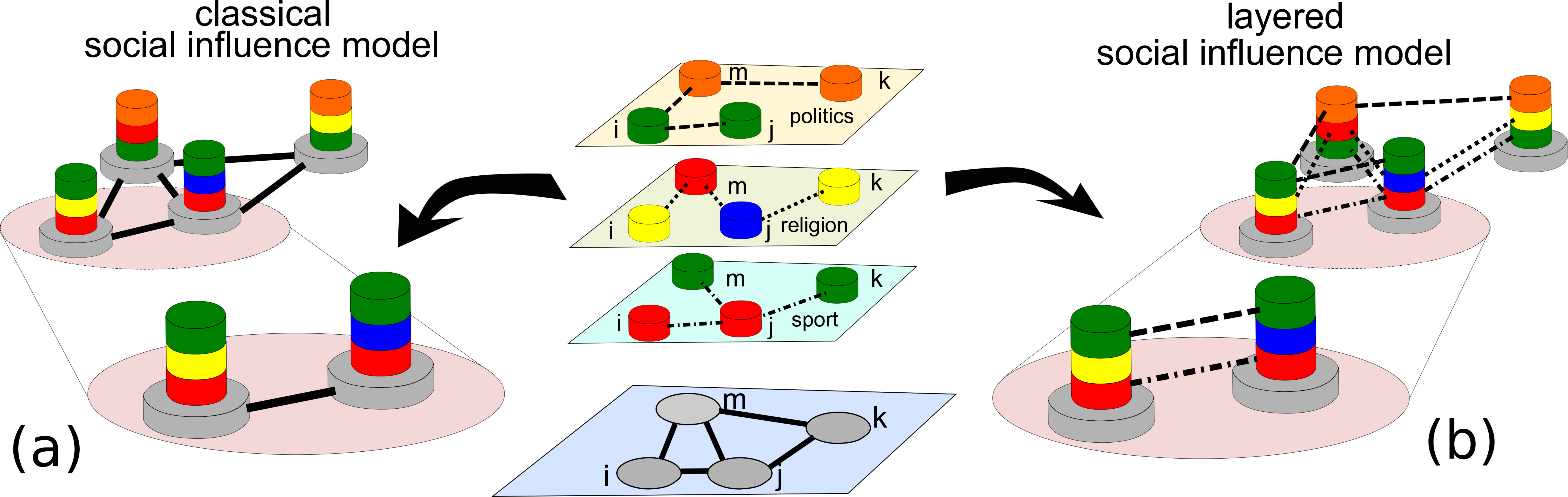}
\caption{\textbf{Social influence is inherently layered.}
Social systems are multiplex networks because individuals tend to differentiate their social contacts according to the different subjects of the interaction. In the classical social influence model, the specific type of each interaction is neglected and all connections are treated together (a). Conversely, we propose here a multilayer cultural influence model in which each interaction can only affect the corresponding cultural features (b). As an example, in the classical model the bond between agents $i$ and $j$ is active at each cultural feature: the two agents have a different religious trait but according to social influence, one of the two agents will eventually absorb the religious trait of the other one. The same bond is instead frozen in the layered model, as imitation can not affect religious beliefs, since the two agents are not linked on the corresponding layer.}
\label{fig:fig1}
\end{center}
\end{figure*}

In this work we propose an alternative mechanism to account for robust multiculturality that naturally brings together social influence and the nature of the social network of interactions with no need of introducing any additional dynamical feature. In real societies
the relationships among individuals are
inherently layered, i.e. individuals tend to interact with different
neighborhoods on different topics. Hence, interactions in social
systems are better described in terms of multiplex
networks~\cite{Boccaletti2014,rev1, battiston17challenges}, i.e., networks composed of many
layers, each one of them describing a particular type of interaction,
with a wide range of overlap between the links at the different
layers as found in multiplex empirical data~\cite{battiston14,szell10,dedomenico15pa,battiston15b, klimek}. We model cultural dissemination associating each cultural feature to a different layer of a multiplex network. Social influence becomes now a layered mechanism in which an agent is allowed to imitate only a subset of all the cultural features of its neighbors, namely those for which there exists a link
on the corresponding interaction layer. The consequences of layered social influence are strongly dependent on the structural overlap among different layers, so that the overlap becomes the control parameter for the nonequilibrium transition in the system. We remark that the multiplex topology has been already successfully applied in other types of social dynamics, such as opinion formation, giving rise to novel critical behavior~\cite{diakonova,battiston15a}.

We find that layered social influence in synthetic and empirical multiplex networks with heterogeneous layers easily leads to a global state of cultural diversity. This state exists for any number of cultural traits provided that the interaction patterns into the population are sufficiently layered, i.e. the value of the structural overlap is below a critical point. We remark that this is a qualitative shift in the behavior of the system, only achievable in multiplex networks, and it is in agreement with empirical evidence of fragmented states even in social contexts with a limited number of available cultural choices $q$. Moreover, unlike the fragmented states found by Axelrod, such new multicultural state is robust against cultural drift, thus providing an explanation for the persistence of multiculturality we experience in real-world society. In addition, we find novel phases of cultural diversity in which a global culture for a number of cultural features coexist with polarization in other features, a situation reminiscent of the so-called chimera states \cite{chimera1,chimera2,chimera3}. More in general, we observe that
different levels of homogeneity may be achieved on different
cultural features, depending on the heterogeneity of the structure of interactions across the layers. Finally, we investigate two layered social networks from the
real-world, showing how considering the nature of the different interactions promotes cultural fragmentation in the population.


\section*{Model}

\begin{figure*}[t]
\begin{center}
\includegraphics[width=6in]{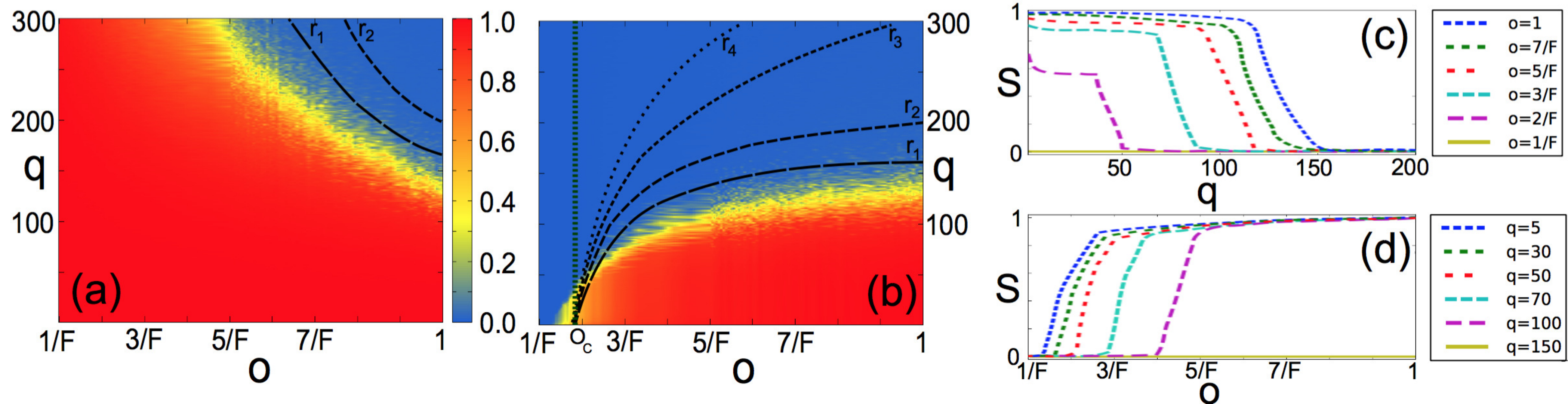}
\caption{\textbf{Modeling classical versus layered social influence.}
  The size of the largest cultural component $S$ in the classical (a)
  and in the layered (b) social influence model is shown as function
  of the number of cultural traits $q$, and for multiplex networks
  with $F=10$ layers and a
  tuneable structural overlap $o$.
  In the case of classical social
  influence, a decrease in $o$ increase the density of the aggregated networks, hence increasing the value
  of $q_c(o)$.
  Conversely, for layered social influence, the critical
  number of cultural traits $q_c(o)$ separating globalisation (red
  region) from multiculturality (blue region) decreases with the
  overlap, and goes to $0$ for $o$ smaller than a critical value. This
  means that, for sufficiently small values of the overlap,
  multiculturality is always achieved independently from the values of
  $q$. Under the presence of drift $r$, the critical overlap sustaining globalisation depends both on $q$ and $r$.
  Noticeably, the polarised phase obtained for $o<o_c(q, r=0) \approx 2/F$ is stable
  under the presence of noise. This is shown in (b), where we plot with different dashes
  the critical lines relative to four increasing values of noise
  $r=r_1,\ldots,r_4$. Conversely, the multicultural states on the classical model are not stable in presence of drift. The two projections of the phase diagram for
  layered social influence are reported in (c) and (d). Transitions
  from globalisation to multiculturality are steep both as a function
  of the number of cultural traits and the overlap.  }
\label{fig:fig2}
\end{center}
\end{figure*}

Let us consider a social systems where agents are involved on interactions on different topics, such as politics, religion and sport, as shown in Fig.~\ref{fig:fig1}. As an example the couple of agents $i$ and $j$ discusses sports and politics, the couple $j$ and $k$ discusses sports and religion, while  $m$ and $k$ can only discuss politics.
In classical models of social influence such as the Axelrod model, the cultural profile of each individual is a vector of different features (sport, religion, politics), shown as a stack of
cylinders, and each feature can take one of a series of different traits, represented as the colors of the cylinders. However, the peculiar nature of each interaction is neglected
and all connections are treated together. Consequently, interactions at any level can potentially impact (and thus modify) any cultural trait,  as in Fig.~\ref{fig:fig1}(a). For instance, the existence of a link between nodes $i$ and $j$, and the different traits of the pair on the feature ``religion'', suggest that this bond is still active. According to social influence, one of the two agents will eventually absorb the religious trait of the other one, making the pair equal
on all the three features. Conversely, we propose a layered cultural influence model where the structure of the relationships pertaining to different social spheres is preserved and explicitly taken into account, assuming that each layered interaction can affect the corresponding cultural features, as shown in Fig.~\ref{fig:fig1}(b). Agents $i$ and $j$ are not linked at the religious layer. As a consequence, since the two individuals already have the same sport and politics traits, the interaction between $i$ and $j$ is frozen. Indeed, the two individuals already reached consensus at all levels where they are linked, and imitation can not concern religious beliefs, since religion is not discussed by them.

We propose in general to take into account the multilayer structure of real
social interactions by modelling layered social influence between pairs
of individuals. Formally, we can describe the specific pattern of social influence at each
cultural level $f$ by mean of a {\em multilayer network}, i.e. a set
of $F$ adjacency matrices $A^{[f]}=\{a_{ij}^{[f]} \}$, $f=1,\ldots,F$,
where $a_{ij}^{[f]}=1$ if agents $i$ and $j$  influence
one another on topic $f$~\cite{Boccaletti2014,battiston14}.

Given two individuals $i$ and $j$, the normalized 
total number of connections
between them is measured by the so-called edge overlap,
$o_{ij}=\frac{1}{F}\sum_{f=1}^F a_{ij}^{[f]}$ which is a number
ranging in the interval $[0,1]$ \cite{battiston14, bianconi13}. We can
then define the {\em structural overlap} $o$ of a multilayer network
as the average number of links among connected individuals:
\begin{equation}
o=\frac{\sum_{i}\sum_{j\neq
    i}o_{ij}\Theta(o_{ij})}{\sum_{i}\sum_{j\neq i}\Theta(o_{ij})},
\end{equation}
where $\Theta(x)$ is the Heaviside function, i.e. $\Theta(x)=1$ if
$x>0$, and $1/F \le o \le 1$. When $o \approx 1$ the layered nature of
social influence will be negligible. Indeed, if the peculiar nature of the links is neglected, the corresponding aggregated network of social interactions (formally defined at the end of this Section) will be similar to the interaction networks relative to each feature.
Conversely, when $o \approx 1/F$ interactions
are extremely diversified across the different topics. As a consequence, the aggregated network will be very different from the interactions occurring at each layer, hinting at possible different outcomes between the layered model and the original model on the aggregated network, to which we refer as the classical case.

The cultural profile
of each agent is described by a feature vector of $F$ integer
variables $\bm s=(s^{[1]} , \ldots, s^{[F]} )$. Each feature $f$, with
$f=1,\ldots,F$, takes one of $q$ possible traits, $s^{[f]}=1,2\ldots,
q$.
Because of the layered structure of interactions, when an agent
interacts with a neighbor it only considers the subset of features
where the two are connected. According to the principle of {\em
  homophily}, the probability of a social interaction between two connected agents
$i$ and $j$ is assumed to be proportional to their layered \textit{cultural
  overlap}, which we define as:
\begin{equation}
   \omega_{ij}=\frac{1}{F}\sum_{f=1}^{F}
  a_{ij}\lay f \delta{\left(s_i^{[f]}, s_j^{[f]}\right)}.
\label{eq:eq1lay}
\end{equation}
where $\delta\left(s_i^{[f]}, s_j^{[f]}\right)$ is the Kronecker's delta
function. Notice that the cultural overlap of two agents is
proportional to the number of shared traits on the layers in which the
two agents are directly connected. When two agents interact, their
cultural profiles are updated according to {\em layered social influence}. In
practice, when agent $i$ interacts with one of its neighbors $j$,
imitation occurs, and $j$ aligns one of its cultural features to that
of $i$, choosing only among cultural features in which $i$ and $j$
interact, i.e., only those features for which a link between $i$ and
$j$ exist in the corresponding layer.

The dynamics of the model proceeds in epochs. During each epoch we
consider each of the $N$ agents in a random order, and perform the
following steps:
\begin{enumerate}
\item{Consider agent $i$, chosen uniformly at random.}
\item{Consider a neighbor $j$ of $i$, chosen uniformly at random
  among all the neighbours of $i$ (nodes with a link to $i$ on at least one layer).}
\item{Let $i$ and $j$ interact with probability $\omega_{ij}$ in
  Eq.~\ref{eq:eq1lay}.}
\item{If the interaction takes place, choose at random a feature $f$
  such that $a_{ij}\lay{f} = 1$ and set $s_j^{[f]} = s_i^{[f]}$.}
\end{enumerate}
The dynamics stops when all pairs of connected agents $i$ and $j$ have
either $\omega_{ij}=0$ or $\omega_{ij}=o_{ij}$.  This implies
the existence and stability of interacting pairs of agents who only
share a limited number of cultural features, a realistic property
observed in real systems but not achievable in the classical model. 

In fact, if we neglect the layered structure of interactions, we are
implicitly assuming that any link between two individuals potentially
allows for an exchange of traits over any feature. In that case, the
structure of interactions in the systems can be described by the
corresponding aggregated network, i.e. the single-layer graph $A = \{a_{ij}\}$
associated to the original multiplex such that $a_{ij}=1$ if $i$ and $j$ have a link in at least one of
the $F$ layers.
We note that the aggregate network does not lead to any discrepancy with the structure of the original multilayer networks for the dynamics of cultural diffusion when $o=1$.


\section*{Results}
In this Section we show how the presence of layered interactions,
which mimic the multilayer structure of real-world
societies~\cite{szell10,battiston15b}, can explain empirical
observations on the presence of multiculturality in real systems. In
order to study the impact of layered interactions on culture
diffusion, we model the structure of a social network by tuning its
level of structural overlap $o$. We considered multilayer networks
with $F$ layers, each of them being an \Erdos-Renyi graph with the
same number of links such that $\langle k\lay{f} \rangle =  \frac{1}{2N}\sum_{i,j \neq i} a_{ij}\lay f = 4$ $\forall f=1, \ldots, F$.

In the case of $F$ identical layers, the
stuctural overlap is maximum and we have $o=1$. Starting with this configuration, we then assign to each edge of each layer a probability $p$ to be
rewired at random. Since each edge is rewired independently, we can
express the structural overlap $o$ as a function of $p$ (see
Methods). In general, the larger the value of $p$, the lower the value
of $o$, with $o=1$ for $p=0$ and $o=1/F$ for $p=1$, and the denser the corresponding aggregated network.

In Fig.~\ref{fig:fig2} we plot the size of the normalised largest cultural
component $S$ as a function of the number of cultural traits $q$ and
for different values of the structural overlap $o$, both for the
classical Axelrod model (a), which runs on the aggregate graph
associated to the multiplex network, and for the layered social
influence model (b). Two individuals are considered to be part of the same cultural component if there is a path from one to the other (paths whose edges lay on different layers are allowed) and if they have equal cultural traits for all $F$ features. In the classical model,
a lower value of overlap by making the aggregated network denser is favouring globalisation, i.e. $S \approx 1$, increasing the critical number of cultural traits $q_c(o)$ at which multiculturality appears, i.e. $S \approx 0$.

\begin{figure*}[tb]
\begin{center}
\includegraphics[width=6in]{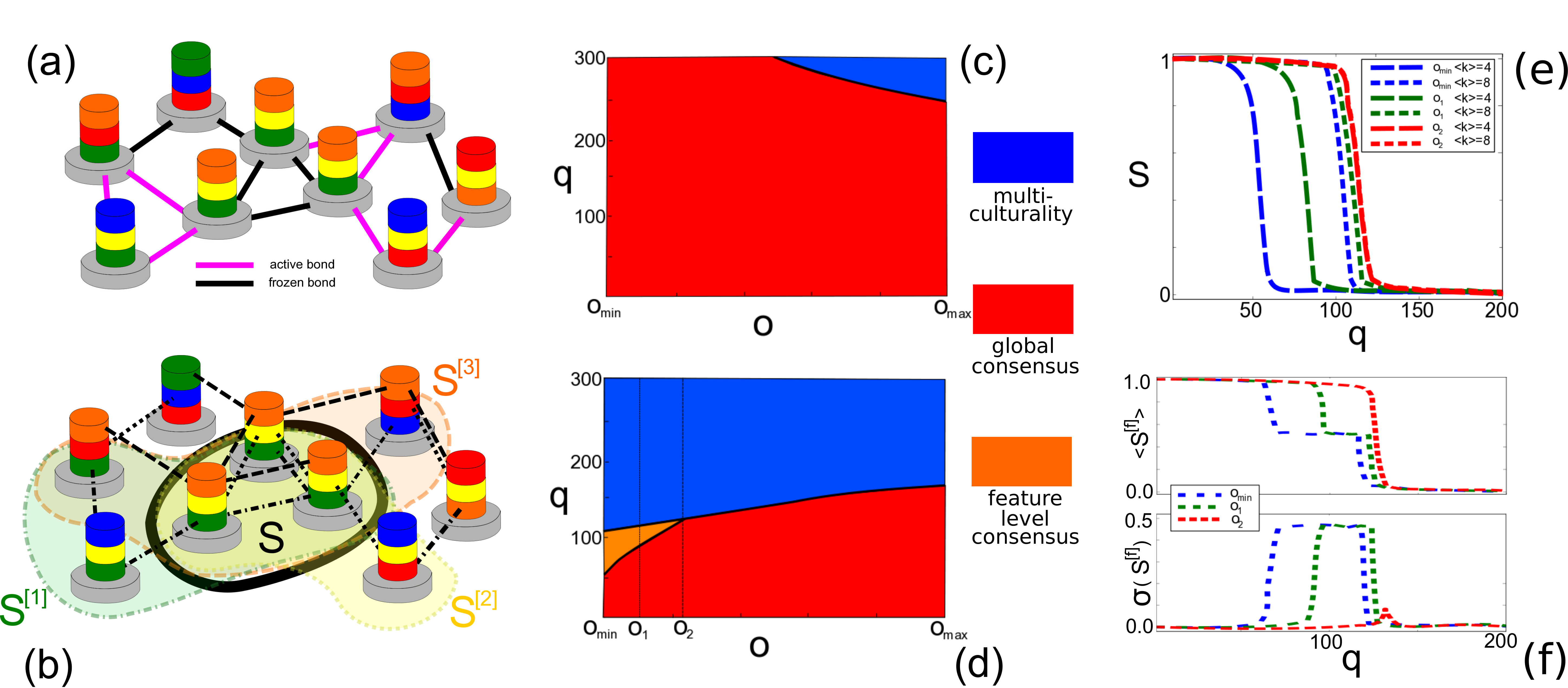}
\caption{\textbf{Global and feature-level consensus.}  A toy social
  system with three cultural features (a, b). With classical social
  influence (a), such system is still far from the
  equilibrium. Conversely, with layered social influence (b), the
  system is already in the absorbing state, and the largest cultural
  components at each layer differ from the global one.  In (c, d) we show
  the phase diagram of the global largest cultural component of the
  system for classical (c) and layered (d) social influence as
  a function of $q$ and for different values of the overlap $o$ in
  networks with $F=10$ and five low-density layers ($\langle
  k \rangle=4$) and five high-density layers ($\langle k
  \rangle=8$). In the classical case, the system reaches a state where
  either globalization (red region) or fragmentation (blue region) is
  achieved for all features. In the layered case, instead, when $o$ is
  smaller than a critical value $o_2$ , it is possible to have
  globalisation in the five high-density layers together with
  multiculturality in the five low-density ones (orange region). In
  (e), for the layered case, we plot the values of $S$ as a function
  of $q$ for one low-density and one high-density layer, and for three
  different values of the overlap, namely $o_{\rm min}$, $o_1$ and
  $o_2$. In (f), for the same values of overlap, we show the average
  and the standard deviation of the largest component $S\lay f$ over
  all layers. }
\label{fig:fig3}
\end{center}
\end{figure*}

For $o=1$ (i.e., when all the layers are identical) the classical and
layered model are undistinguishable, and multiculturality can only be
achieved for large values of $q$. For instance, in the population considered in Fig.~\ref{fig:fig2}, multiculturality can
be achieved only for $q>140$. In the layered social influence model, where an agent can absorb the cultural trait of his neighbor only if they are connected
instead, $q_c(o)$ becomes smaller as $o$ decreases, until it vanishes at
critical value $o_c(q)>0$ $\forall q \ge 2$, whose exact value depends on $q$, as shown in Fig.~\ref{fig:fig2}(c). This implies
that for low structural overlap globalisation is not achievable, and the system always has a qualitative different behavior, always converging towards a
multicultural state
independently of the number of cultural traits $q \ge 2$. We note that in the classical model with $o=1$ multicultural states can only be obtained for an unrealistically high number of cultural traits $q \approx 125$.

A symmetric situation is observed if we study the system as a function of $o$, as
shown in Fig.~\ref{fig:fig2}(d), where it is evident that when $q$ is
small fragmentation can be obtained only for very low values of
structural overlap. As $q$ increases, the globalised phase shrinks and
the critical value $o_c$ increases. For $q>140$, finally, the system
is always in a fragmented state, even in the case of maximal
structural overlap. 

An important feature of  social systems is the presence of cultural
drift, i.e., the occasional spontaneous mutation of a cultural trait
of an agent. Such phenomenon can be modelled as a noise of constant rate
$r$, acting on the system on longer time-scales compared to the one
that regulate the imitation and interactions among individuals. In the
classical Axelrod model on finite populations the multicultural absorbing states (blue
region in Fig.~\ref{fig:fig2}(a)) are known to be metastable and
fragmentation is destroyed if spontaneous mutation is allowed
\cite{klemm03, klemm05}, still leaving unanswered how it is possible to explain the persistence of multiculturality in realistic societies.

In Fig.~\ref{fig:fig2}(b) we show with
different dashed lines the position of the critical line separating
globalisation from fragmentation under four increasing values of
noise, namely $r_1=3 \times 10^{-5}$, $r_2=5 \times 10^{-5}$, $r_3=6
\times 10^{-5}$ and $r_4=7 \times 10^{-5}$. We note that, given a number of cultural traits $q$, the critical value of overlap needed to sustain globalisation depends on the amount of noise in the system, i.e. $o_c = o_c(q,r)$. In analogy to the
classical case, for $o>o_c(q,r)$ the multicultural region appearing at high
values of $q$ is unstable under cultural drift. Conversely, the
polarised phase obtained for $o<o_c(q,r)$ is not affected by the presence
of noise and allows to explain the persistence of multicultural states
that we observe in real-world societies. Noticeably, in the limit $q \to 2$ the value of $o_c(q,r)$ appears to be independent of $r$ and approaching the value of $2/F$.  
Conversely, in the classical model shown in Fig.~\ref{fig:fig2}(a) already all multicultural states for $q<200$ are destroyed already for a drift as small as $r_2=5 \times 10^{-5}$, and all multicultural states for $q<300$ are lost for $r_3=6
\times 10^{-5}$, no matter the value of the structural overlap. We remark that the instability of multicultural states is meant under the introduction of a moderate level of drift. Conversely, if drift is too high, the system enters a noisy dynamical state, describing an unrealistic society where spontaneous mutation accounts for most of the cultural chances and the effect of social influence is widely neglected.



Individuals might reach consensus on some topics and disagree on some other ones. To this scope, we introduce now the largest topical cultural component $S\lay f$. Two individuals $i$ and $j$ are considered to be part of the same topical component $f$ if there is a path connecting them at layer $f$, and share the same cultural trait on feature $f$, i.e. $s_{f(i)} = s_{f(j)}$. We note that when $S=1$, the condition that $S\lay f=1$ $\forall f$ is automatically satisfied in both the classical and the layered model (assuming that each layer is connected). A remarkable property of layered social influence is that it allows for
different levels of consensus on single topics to coexist, possibly
resulting in the emergence of globalisation on some topics, i.e. $S\lay f \approx 1$ for some features $f$, and the
persistence of fragmentation in some other ones, $S\lay {\tilde f} \approx 0$ for some other $\tilde f$. We call this regime \textit{feature-level consensus}, or
\textit{topical consensus}.  We remark that this realistic phenomenon
can not be achieved with the classical version of the model, where at
the absorbing state the largest component on each layer is always as large as
the one computed taking into account all topics simultaneously.

In Fig.~\ref{fig:fig3} we show how feature-level consensus may be
possible only as an effect of the difference in the number of social
interactions related to the different features, i.e. different layers'
densities. Let us consider two sets of layers. The first set is made of
 five equal layers with average degree $\langle k\lay f
\rangle=8$, whereas the second is made of five equal
layers with average degree $\langle k\lay f \rangle=4$. When all edges of the layers in the second set also belong to those in the first set, the
structural overlap in the system is maximised, $o=o_{\rm max}$.
Conversely, when no edges in the low-density layers are
also present in the high-density ones the overall will be minimum, $o=o_{\rm
  min}$, and the system behaves as two isolated five-layer
classical models with different average degree.  Hence, because of the
different density, they will have a different critical values $q_c$
for transitioning from globalisation to multiculturality. Intermediate configurations when $o_{\rm min}<o<o_{\rm max}$ are also considered.

For $o=o_{\rm min}$, the critical values separating globalisations from multiculturality for the low-density and high-density layers are different.
While both sets of layers are globalised for $q<60$ and fragmented for $q>120$, in the range $60<q<120$ globalisation is achieved only in the high-density layers, for which $S\lay f \approx 1$, and the low-density ones are instead polarised, $S\lay f \approx 0$.
Strikingly, such property of the system is
not peculiar of the case $o=o_{\rm min}$, but it is preserved for a
finite range of values of structural overlap of to a critical value
$o_2$. Such region is highlighted in orange in the diagram shown in
Fig.\ref{fig:fig3}(d), and separates the globalised region (red)
from the fragmented one (blue) when the value of overlap is
sufficiently low.

The value of $S\lay f$ for low-density and a high-density layers, as well as the average value of
the largest cultural components $\langle S\lay f \rangle$ and its
standard deviation, are
also shown in Fig.~\ref{fig:fig3}(e, f),. The standard deviation $\sigma(S\lay f)$,
approximately $0$ for both the globalised and fragmented regions,
takes values close to $0.5$ in the mixed orange region, where
consensus only exists on some layers.

We remark that higher heterogeneity in the structure of the layers produces even richer and more diverse patterns of consensus across the different cultural features.
A typical case is when the activity of the nodes on the layers is heterogenous~\cite{nicosiacorr}, i.e. when not all
agents are involved in discussion with other individuals on all
topics, preventing the spread of a cultural traits across the whole
population for some given features. This is the case of many real-world systems, where the structure of
interactions is often inherently layered and diverse levels of activity occur at the different layers. As a test-case, we study the
dynamics of our model with layered social influence on two multilayer
social systems, namely the network of Indonesian
terrorists~\cite{battiston14} and the collaboration network of the
Pierre-Auger observatory~\cite{dedomenico15pa} (see Methods).

\begin{figure}[!t]
\begin{center}
\includegraphics[width=3in]{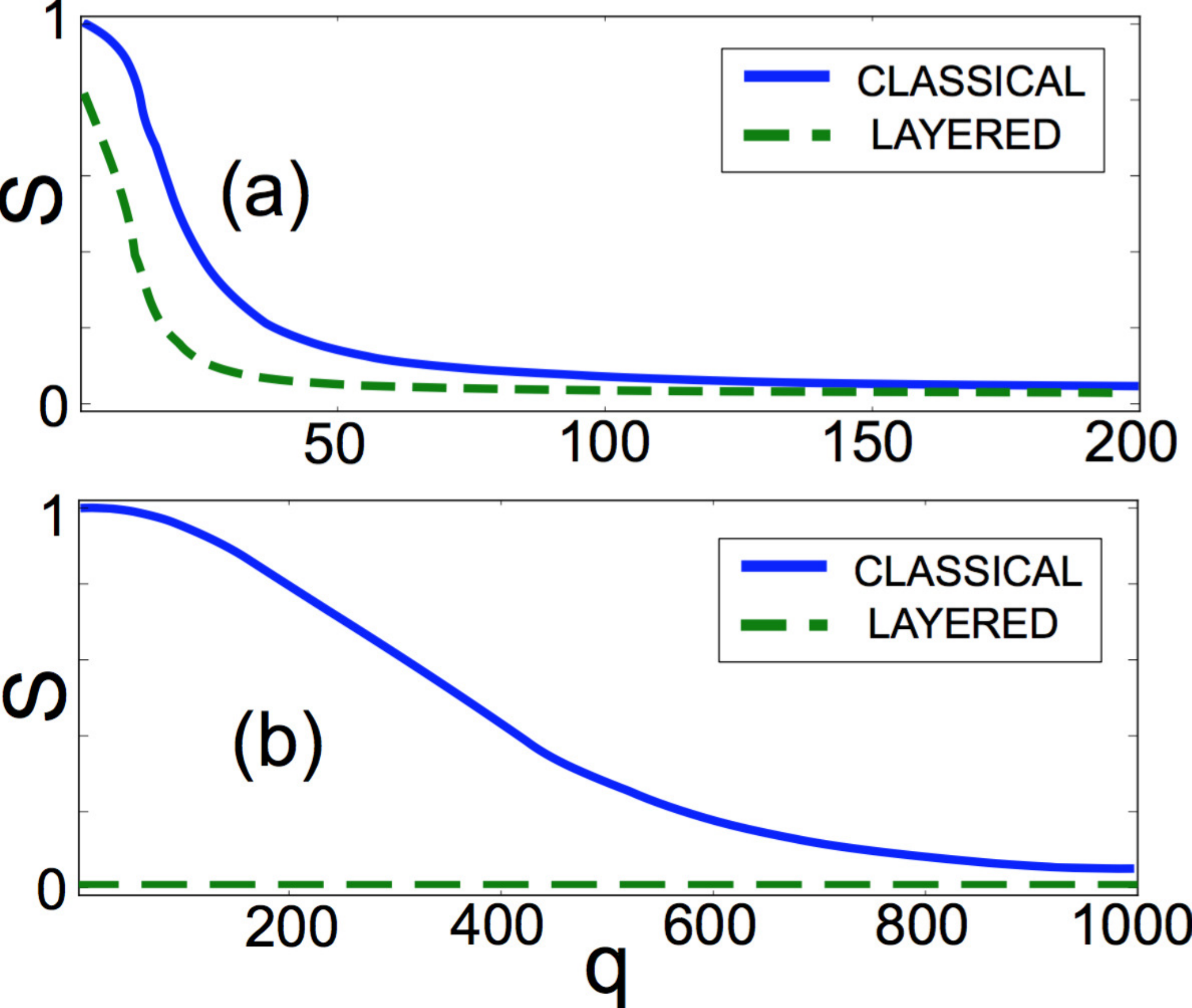}
\caption{\textbf{Multiculturality in empirical multilayer social networks.}  The largest
  cultural component $S$ for classical (solid lines) and layered
  social influence (dashed lines) is reported as a function of $q$ for
  two real-world multiplex social systems, respectively the network of
  Indonesian terrorists in panel (a) and the collaboration network of
  the Pierre Augier observatory in panel (b). In the first case the
  outcome of the classical and layered model are similar, with layered
  social influence only producing a decrease in the value of
  $q_c$. This is mainly due to the relatively large amount of
  structural overlap in the terrorists network ($o=0.48$) and low heterogeneity in the activity of nodes. Conversely,
  the low value of redundancy in the Pierre Augier collaboration
  network ($o=0.07)$, and the heterogeneity in the density and
  activity of the layers, produce qualitatively different results: the
  layered model is always fragmented, even at $q=2$, whereas the
  classical model produces a largest component of order $1/N$ only for
  $q>1000$.}
\label{fig:fig4cap}
\end{center}
\end{figure}

In Fig.\ref{fig:fig4cap} we plot for both systems, as a function of
$q$, the size of the largest cultural components found in the model
with layered social influence (dashed lines) compared with those
obtained with the classical model simulated on the corresponding
aggregated networks (solid lines). For the terrorists network (panel
(a)), a giant component of the order of the number of nodes appear for
small $q$ even in the layered case, while for $q\approx 20$
fragmentation appears on all features. The classical model predicts
instead $q_c\approx 50$. For the Pierre-Auger collaboration network
(panel (b)), in the classical model we have $S\approx 1/N$ only for
$q>1000$. Conversely, the layered model predicts multiculturality
already for $q=2$, producing a qualitative different result. Such striking difference between the layered and classical
dynamics is due to the relatively low level of structural overlap and
to the heterogeneity in the activity and density of the layers of the
multiplex. Such qualitative different behaviors highlights the
importance of the underlying structure of social interactions to
understand cultural patterns in real-world societies.

\section*{Discussion}

Reconciling the phenomenon of social influence and imitation with the
empirical evidence of multiculturality at a global scale is a
long-debated problem. The model for the
dissemination of culture introduced by Axelrod in $1997$ \cite{axelrod97} has proven to
be able to produce socially diversified states in spite of the
existence of a locally polarising rule. However, such
multicultural states occur only for a very high number of
cultural traits $q$ and are not robust against cultural drift. In spite of several proposals \cite{centola07,Flache2006,Flache2011,Flache2011b,Mas2014}, explaining the emergence and persistence of
multicultural states that we experience in our life still appears as
an open problem. In our work we considered agents having different
interaction patterns according to the different cultural
features. This is a property of many social systems, where
individuals usually choose the recipients of their discussion
according to the content of their messages and vice versa. By
constraining the imitation of the cultural traits to the cultural
features where two individuals are actually linked, we naturally
introduce the notion of layered social influence, inherently connected
to the average structural overlap $o$ of a social system. Such
definition implies the existence of stable states where connected
agents might not have completely equal or different cultural profiles,
but instead share just a limited number of features.



A main finding is that a robust multicultural regime emerges as
a natural consequence of taking into account the actual structure of
social interactions, which are inherently layered. Interestingly, when
the structural overlap among the layers is small, i.e. when
connections are not too much redundant across layers, only
multicultural states will be allowed, independently of the actual
number of possible cultural traits, suggesting a qualitative shift in the behavior of the system.  This finding is also in agreement with the empirical evidence of socially
fragmented societies even under a limited number of cultural choices. Moreover, differently
from the multicultural regime predicted when the layering of social influence is not taken into account, and achievable only for a
large number of traits, such novel multicultural regime is
robust to the presence of cultural drift, providing a potential
explanation to the persistence of global diversity over time.

Another interesting aspect of layered social influence is the ability
to explain the existence of a novel regime where globalisation appears
only on a limited number of cultural features, while the remaining
ones are characterised by multiculturality. The coexistence of
different levels of multiculturality in different aspects of a
cultural profile is in fact another typical property observed in human
societies.

In conclusion, the model of layered social influence
proposed in this paper, provides a mechanism leading to robust multiculturality, reveiling the important role played by a
multilayer organisation of social interactions in avoiding
globalisation, and pointing out the effect of heterogeneity of layer
densities in the emergence of partially globalised regimes. While previous proposals to account for robust multiculturality invoke different forms of interactions among agents in an aggregated social network, we provide here an alternative mechanism not based in the form of the interaction, but on the structural properties of the social network. From this new perspective, the relevant control parameter of polarization-globalization transitions is no longer the number of cultural traits per feature, but a structural property, namely the overlap parameter accounting for the heterogeneity of social links among the different layers of the social network. The relevance of the structural properties of multilayared social interactions, illustrated here for cultural dissemination, is far reaching, and should be taken into account to reconsider a number of results based on models of social interactions among agents in aggregated networks.

\section*{Methods}
\textbf{Constructing synthetic multilayer social networks}. In this
section we focus on the model to produce synthetic multilayer networks
with different values of structural overlap $o$. Let us consider a
multiplex network with $F$ layers, each with the same number of links
$K^{[f]}$, so that we have $K^{[f]}F$ links in total. We then consider the
aggregated network obtained by collapsing all the layers. If all the
layers are identical, the aggregated graph will have $K=K^{[f]}$
edges. Hence $o=\frac{1}{F}\frac{KF}{K}=1$ for $p=0$. When $p\neq0$
and links are rewired independently at each layer, the number of edges
in the aggregated networks increases. If we neglect the probability
that two links of different layers are rewired in the same position,
the number of edges in the aggregated network becomes equal to
\begin{equation}
\small
K=\sum_{m=1}^F \binom {F} {m}(1-p)^m p^{F-m}K^{[f]} + \sum_{m=0}^F \binom {F} {m}(1-p)^m p^{F-m}K^{[f]}(F-m)
\end{equation}
However the total number of connections in the multiplex remains
$K^{[f]}F$. Hence
\begin{equation}
o\approx \frac{1}{\sum_{m=0}^F \binom {F} {m}(1-p)^m p^{F-m}(1-\delta_{0,m}+F-m)},
\end{equation}
higher values of rewiring correspond to lower values of overlap and
vice versa, with the limiting cases of $o=1$ for $p=0$ and $o=1/F$ for
$p=1$. Given the average degree $\langle k^{[f]} \rangle$, the average
number of neighbours $\langle k \rangle$ in the aggregated network can
be obtained as $\langle k \rangle=\langle k^{[f]} \rangle/o$. Hence,
the higher the $p$, the lower the $o$ and the higher the density of
the aggregated network.  The results shown in Fig.~\ref{fig:fig2} have
been obtained for a system with $N=25^2$ agents. Each interaction
layer $f$ has average degree $\langle k\lay f
\rangle=\frac{1}{2N}\sum_{i,j \neq i} a_{ij}\lay f=4$.
In Fig.\ref{fig:fig3} we consider layers with different average degrees $\langle
k\lay f \rangle=\frac{1}{2N}\sum_{i,j \neq i} a_{ij}\lay f$. Five of
them are equal networks with $\langle k\lay f \rangle=4$, the other
five are equal networks with $\langle k\lay f \rangle=8$. We consider
different combinations of the two sets of layers such in a way to span
the range of values of structural overlap $o \in [o_{\rm min}, o_{\rm
    max}]$. When $o=o_{\rm max}$, all edges in the emptier layers are
also present in the denser ones. Conversely, when $o=o_{\rm min}$ no
edge of the emptier layer is shared with the denser ones, and the
system behaves as two five-layer decoupled classical models.

\textbf{Datasets}.  In Fig.\ref{fig:fig4cap} we apply the
classical and layered social influence models to two real-world multilayer
networks with layered interaction patterns and different activities of the node and layers. A node $i$ is defined as active on layer $f$, i.e. $b_i\lay f=1$, if $k_i\lay{f}>0$. Otherwise it is consider inactive, i.e. $b_i\lay f=1$~\cite{nicosiacorr}. The first dataset is the network of Indonesian terrorists
presented in~\cite{battiston14}, which has $F=3$ layers, corresponding
to trust, common operations and exchanged communication across $78$
individuals, most of which are active on all layers, and has a
structural overlap $o \approx 0.4817$, with moderate difference in the layers' activity.  The second data set is the
multi-layer collaboration network of the Pierre-Auger
observatory~\cite{dedomenico15pa}, which consists of $F=16$ layers. To
allow, at least potentially, complete globalisation in the system, we
selected the subset of individuals who belong to the giant component
of the aggregated network. Such system includes $475$ individuals. The
system has a structural overlap of $o \approx 0.069$ and high
heterogeneity in the activity of its members.


\section*{Acknowledgments}
The authors acknowledge the support of the EU Project LASAGNE,
Contract no. 318132 (STREP). V.L. also acknowledges support from EPSRC projects EP/N013492/1. M.S.M. also acknowledges financial support from FEDER (EU) and MINECO (Spain) under Grant ESOTECOS FIS2015-63628-C2-R.

\section*{Author contributions}
F.B., V.N. V.L. and M.S.M. designed research, F.B. and V.N. performed research, F.B., V.N. V.L. and M.S.M. analysed data, F.B., V.N. V.L. and M.S.M. wrote the paper.

\section*{Competing financial interests}
The authors declare no competing financial interests.

\end{document}